\documentclass[10pt]{wlscirep}
\usepackage[utf8]{inputenc}
\usepackage[T1]{fontenc}
\usepackage{setspace}
\usepackage{multirow}
\usepackage{phonetic}

\usepackage{kotex}
\usepackage{array}
\usepackage{lipsum}
\usepackage{booktabs}
\usepackage{setspace}
\usepackage{amsmath}
\usepackage{hyperref}
\usepackage{lineno}
\usepackage{xcolor}
\usepackage{xurl}
\usepackage{tipa}

\usepackage{fancyhdr}
\usepackage{lastpage}

\fancypagestyle{plain}{%
  \fancyhf{}
  \fancyfoot[R]{\thepage/\pageref{LastPage}}

}

\usepackage{caption}
\newcommand\blfootnote[1]{%
  \begingroup
  \renewcommand\thefootnote{}\footnote{#1}%
  \addtocounter{footnote}{-1}%
  \endgroup
}

\title{Throat and acoustic paired speech dataset for deep learning-based speech enhancement}

\author[1,$\dag$]{Yunsik Kim}
\author[1,$\dag$]{Yonghun Song}
\author[1,2,3,*]{Yoonyoung Chung}
\affil[1]{Department of Electrical Engineering, Pohang University of Science and Technology (POSTECH), 77 Cheongam-ro Nam-gu, Pohang, Gyeongbuk, 37673, Korea}
\affil[2]{Graduate School of Semiconductor Technology, Pohang University of Science and Technology (POSTECH), 77 Cheongam-ro Nam-gu, Pohang, Gyeongbuk, 37673, Korea }
\affil[3]{Intus Co. Ltd., 87 Cheongam-ro Nam-gu, Pohang, Gyeongbuk, 37673, Korea}

\affil[*]{Corresponding author: Yoonyoung Chung (ychung@postech.ac.kr)}

\affil[$\dag$]{These authors contributed equally to this work}

\begin{abstract}
In high-noise environments such as factories, subways, and busy streets, capturing clear speech is challenging. Throat microphones can offer a solution because of their inherent noise-suppression capabilities; however, the passage of sound waves through skin and tissue attenuates high-frequency information, reducing speech clarity. Recent deep learning approaches have shown promise in enhancing throat microphone recordings, but further progress is constrained by the lack of a standard dataset. Here, we introduce the Throat and Acoustic Paired Speech (TAPS) dataset, a collection of paired utterances recorded from 60 native Korean speakers using throat and acoustic microphones. Furthermore, an optimal alignment approach was developed and applied to address the inherent signal mismatch between the two microphones. We tested three baseline deep learning models on the TAPS dataset and found mapping-based approaches to be superior for improving speech quality and restoring content. These findings demonstrate the TAPS dataset's utility for speech enhancement tasks and support its potential as a standard resource for advancing research in throat microphone-based applications. 
\end{abstract}

\begin{document}
\maketitle
\blfootnote{This is a preprint of an article published in \textit{Scientific Data}. The final authenticated version is available online at: \href{https://doi.org/10.1038/s41597-026-07268-2}{https://doi.org/10.1038/s41597-026-07268-2}}

\flushbottom

\section*{Background \& Summary}
Capturing high-quality speech through acoustic microphones is often limited in real-world environments with substantial background noise. Body-conducted microphones (BCMs) with noise-suppressing capabilities, such as throat and in-ear devices, have been used to enable effective communication in noisy environments. These devices capture speech information transmitted from the vocal cords and vocal tract to the skin surface. Researchers have proposed various BCMs using piezoelectric\cite{lee2013highly, dagdeviren2014conformable, park2015fingertip}, piezoresistive\cite{kim2016body, park2016dramatically, qiu2015ultrafast}, piezo-capacitive\cite{zang2015flexible, jin2017ultrasensitive, lee2019ultrathin}, triboelectric\cite{fan2015ultrathin, yang2015eardrum, kang2018transparent}, and electromagnetic\cite{zhao2020fully, gao2022comparison, zheng2022dual} materials, as well as commercial accelerometers\cite{song2025multimodal}. These microphones exhibited high sensitivity to voice-related vibration signals and were designed with soft form factors that fit comfortably on curved skin surfaces, which makes them suitable for wearable communication devices. However, when a speech signal from the vocal tract transmits through the skin and muscles, it experiences the low-pass effect that attenuates high-frequency components\cite{shin2012survey}. In particular, BCM recordings reduce high-frequency aperiodic components critical for sibilants (e.g., /s/, /z/, /\textipa{\textesh}/) and aspiration (e.g., /h/)\cite{AckerMills2005}. This attenuation degrades timbre and reduces speech intelligibility by diminishing consonant distinctiveness\cite{AckerMills2005, tran2013effect}. Moreover, certain phonemes produced within the oral cavity rather than the vocal cords—especially voiceless obstruents such as voiceless fricatives, plosives, and affricates—are not effectively captured by BCMs\cite{toda2012statistical}. Additionally, inappropriate placement of the sensor apart from the vocal cord can further degrade sound quality\cite{mcbride2011effect, song2021study}. Therefore, developing effective speech enhancement techniques is crucial when utilizing BCMs for speech measurement.

In early studies for improving speech information from BCMs, statistical models like linear prediction\cite{vu2007blind, rahman2017lp} and Gaussian mixture models\cite{nakagiri2006improving, toda2012statistical, turan2015source} were used. These models are based on the source-filter model, which represents speech as an excitation and a spectral envelope filter. The excitation source is assumed to be the same for the speech captured by the BCM and the corresponding acoustic microphone. Consequently, the speech enhancement task is simplified to modifying vocal tract filter characteristics, such as line spectral frequency \cite{toda2012statistical, vu2007blind} and Mel cepstrum coefficients \cite{nakagiri2006improving, turan2015source}. However, mutual independence of the source and the filter is not strictly guaranteed, and low-dimensional spectral envelopes are insufficient to characterize speech well, which results in poor speech enhancement performance \cite{gao2022comparison}.

Recent advances in deep learning have enabled significant progress in body-conducted speech enhancement, as these methods can model high-dimensional speech features. Researchers have developed various enhancement methodologies, including deep denoising autoencoders\cite{huang2017wearable, liu2018bone}, bidirectional long short-term memory (BLSTM)\cite{zheng2018novel, gao2022comparison}, and dual-path transformers (DPT)\cite{zheng2022dual}. For instance, enhancing BCM speech with denoising autoencoders has demonstrated increased speech quality and reduced error rate in automatic speech recognition (ASR) systems. Furthermore, BLSTM and DPT-based models can generate missing speech information by leveraging the temporal dependencies of features, thereby improving representative metrics such as the Perceptual Evaluation of Speech Quality (PESQ) and Short-Time Objective Intelligibility (STOI).

Despite these advances, further progress is hindered by the absence of a large-scale, standardized, and meticulously aligned dataset for training and evaluating BCM-based speech enhancement models. Consequently, prior studies have often relied on smaller, in-house datasets collected under heterogeneous recording setups (e.g., sensor type/placement and recording configurations)\cite{zheng2018novel, zheng2022dual, huang2017wearable, liu2018bone, gao2022comparison}. Because these datasets are rarely public or standardized, direct quantitative comparisons across studies are difficult, hindering reproducibility and limiting generalizability. To address this gap, we publicly release the Throat and Acoustic Paired Speech (TAPS) dataset as a standardized resource for training and evaluating BCM-based speech enhancement models. Several BCM datasets have been introduced to facilitate the development and evaluation of body-conducted speech enhancement models. For example, the ESMB (repository: \href{https://github.com/elevoctech/ESMB-corpus}{https://github.com/elevoctech/ESMB-corpus}) and ABCS\cite{ABCS} (repository: \href{https://github.com/wangmou21/abcs}{https://github.com/wangmou21/abcs}) datasets contain Chinese speech recorded using an ear-canal BCM. The Vibravox\cite{Vibravox, Vibravox_data} dataset consists of French speech recorded using five BCMs placed at different locations (e.g., forehead and temple). However, these datasets do not systematically address the temporal misalignment between signals acquired from BCMs and acoustic microphones. Ensuring synchronization between modalities is critical when constructing datasets from multiple sensors, as it directly affects the quality and reliability of the data for deep learning training. Moreover, to the best of our knowledge, no publicly available dataset includes speech recorded using a BCM placed on the supraglottic area of the neck. This region is anatomically close to the vocal cords and advantageous for capturing vibrations with accurate harmonic content and high signal-to-noise ratio (SNR) \cite{song2021study}. The TAPS dataset directly addresses these limitations by providing meticulously aligned paired recordings with a throat microphone specifically positioned on this advantageous supraglottic area.

This paper is organized into two main parts. The first part details the use of a throat microphone as BCM and presents a standard pipeline for constructing the TAPS dataset, specifically tailored for training throat microphone speech enhancement (TMSE) models. We developed a hardware system for simultaneously recording speech from throat and acoustic microphones and collected the data. We examined multiple factors that affect signal mismatches (i.e., temporal misalignments) between the two modalities, including speaker variability, sentence content, and the distance between the speaker's lips and the acoustic microphone. Based on this analysis, we applied an optimized alignment procedure to generate high-quality paired data suitable for training TMSE models. The TAPS dataset is divided into three subsets: \texttt{train}, \texttt{dev}, and \texttt{test}. The \texttt{train} set comprises 10.2 hours of audio and contains 4,000 paired utterances recorded from 40 native Korean speakers. The \texttt{dev} set includes 2.5 hours of audio from 1,000 paired utterances by another 10 native Korean speakers and is used for hyperparameter tuning. The \texttt{test} set consists of 2.6 hours of audio and includes 1,000 paired utterances from another 10 native Korean speakers. In the second part, we report the results of training TMSE models using the TAPS dataset. The experiments were conducted using three models: two-stage transformer neural network (TSTNN)\cite{wang2021tstnn}, Demucs\cite{defossez2020real}, and speech enhancement convolution-augmented transformer (SE-conformer)\cite{kim21seconformer}. Mapping-based models such as Demucs and SE-conformer achieved notable improvements in speech quality and content restoration. We also explored the impact of signal mismatch between throat and acoustic microphones on the performance of deep learning models. The TAPS dataset will facilitate the practical application of throat microphones in extreme noise environments through robust speech enhancement models. It can also contribute to the development of broader speech-related technologies, such as ASR and silent speech interfaces. Furthermore, by establishing a standard dataset collection framework, this methodology can be extended to various languages and applied in diverse domains such as assistive communication \cite{Speech2024Kwon} and voice-based human–computer interaction \cite{Improving2009Erzin}.

\section*{Methods}
\subsection*{Speaker information}
The TAPS dataset was constructed from 60 native Korean speakers, whose demographic details are summarized in Table \hyperref[table:1]{1}. Participants were recruited via on-campus bulletin board advertisements at Pohang University of Science and Technology (POSTECH). We aimed for balanced gender representation and imposed no age restrictions. The mean age of the speakers was 27.1 years (standard deviation: 6.23 years), and none reported a history of vocal disorders. The \texttt{train} set includes 40 speakers (20 women, 20 men). The \texttt{dev} and \texttt{test} sets each consist of 10 speakers (5 women, 5 men per set), distinct from those in the \texttt{train} set and from each other.

\begin{table}
\centering
\caption{Summary of dataset characteristics.}
\label{table:1}
\begin{tabular}{c|ccc}
\toprule
Dataset type & Train & Dev & Test \\
\hline
\begin{tabular}[c]{@{}c@{}}Number of speakers\end{tabular} & 40 & 10 & 10\\
\begin{tabular}[c]{@{}c@{}}Number of male / female speakers\end{tabular} & 20 / 20 & 5 / 5 & 5 / 5\\
\begin{tabular}[c]{@{}c@{}}Mean / standard deviation of the speakers' age\end{tabular} & 28.5 / 7.3 & 25.6 / 3.0 & 26.2 / 1.4 \\
\begin{tabular}[c]{@{}c@{}}Number of utterances\end{tabular} & 4,000 & 1,000 & 1,000 \\
\begin{tabular}[c]{@{}c@{}}Total length of utterances (h)\end{tabular} & 10.2 & 2.5 & 2.6 \\
\begin{tabular}[c]{@{}c@{}}Max / average / min length of utterances (s)\end{tabular} & 26.3 / 9.1 / 3.2 & 17.9 / 9.0 / 3.3 & 16.6 / 9.3 / 4.2 \\
\bottomrule
\end{tabular}
\end{table}

\subsection*{Recording hardware configuration}
We developed a custom-built system to simultaneously record speech from throat and acoustic microphones (Fig. \hyperref[fig:1]{1}a, b).

\subsection*{Microphone specifications}
The throat microphone consisted of a micro-electromechanical systems (MEMS) inertial measurement unit (IMU; TDK IIM-42652) configured as an accelerometer to capture vibration signals from the neck skin surface above the thyroid cartilage. The accelerometer was configured with 8 kHz sampling rate, 16-bit resolution, and $\pm$4g dynamic range, enabling the accurate capture of a wide range of signal amplitudes. We recorded only the z-axis accelerometer channel; the gyroscope and the remaining acceleration axes were not recorded. This accelerometer, mounted on a 1.7 mm thick FR-4 printed circuit board, formed the throat microphone assembly, weighing only 0.76 g. It was secured with an adjustable thin strap designed for close conformity to the neck surface without restricting vocalization or movement. For acoustic speech signal acquisition, a MEMS acoustic microphone (CUI Devices, CMM-4030D-261) operating at a 16 kHz sampling rate and 24-bit resolution was used.

\subsection*{Signal processing and transmission}
The measured vibration and speech signals were transmitted to the microcontroller unit (MCU, STMicroelectronics, STM32F301C8T6TR), which was integrated into the peripheral board, via a serial peripheral interface (SPI) and an integrated interchip sound (I²S) interface, respectively. The inherent delay between the two signals at the MCU ranged from 0.75 to 0.88 ms; this was subsequently synchronized during post-processing. The MCU encoded the paired signals in HEX format and transmitted them to a connected laptop via RS-232 communication, enabling real-time recording.

\begin{figure}
\centering
\includegraphics[width=\linewidth]{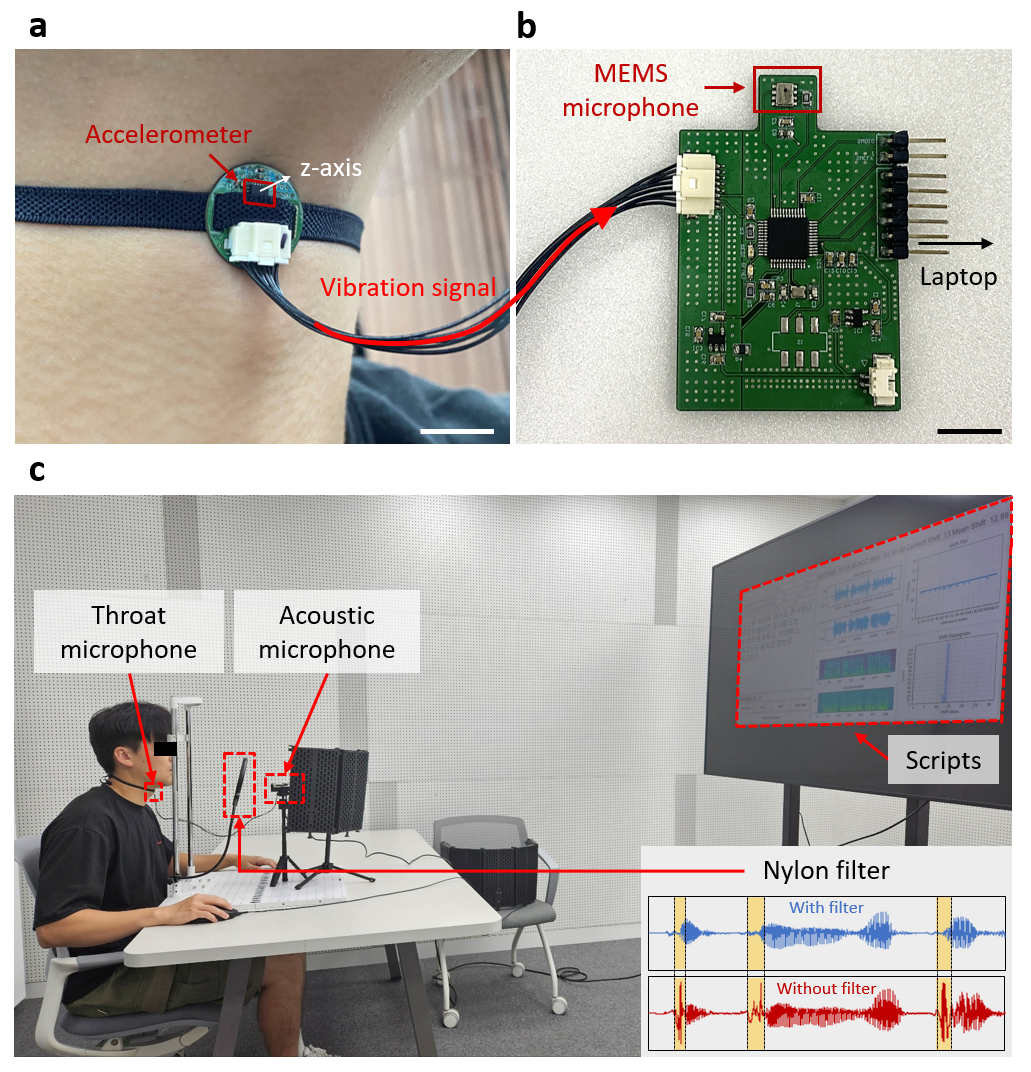}
\caption{\textbf{Experimental setup for simultaneous voice measurement using both throat and acoustic microphones.} (a) The throat microphone contained an accelerometer to capture vibrations from the neck skin. Scale bar: 10 mm. (b) The peripheral board was composed of an acoustic microphone and a microcontroller unit. Scale bar: 10 mm. (c) Photograph of the setup used for voice recordings with 60 speakers in an acoustically treated recording room. The throat microphone was attached to the skin approximately 1 cm above the vocal cords, and the acoustic microphone was placed 30 cm in front of the lips. Additionally, a reflection filter was used to reduce ambient noise, and a forehead rest was provided to help speakers maintain a consistent head position. Speakers read 100 randomly selected sentences from a Korean newspaper corpus (\href{https://corpus.korean.go.kr}{https://corpus.korean.go.kr}).}
\label{fig:1}
\end{figure}

\begin{figure}
\centering
\includegraphics[width=\linewidth]{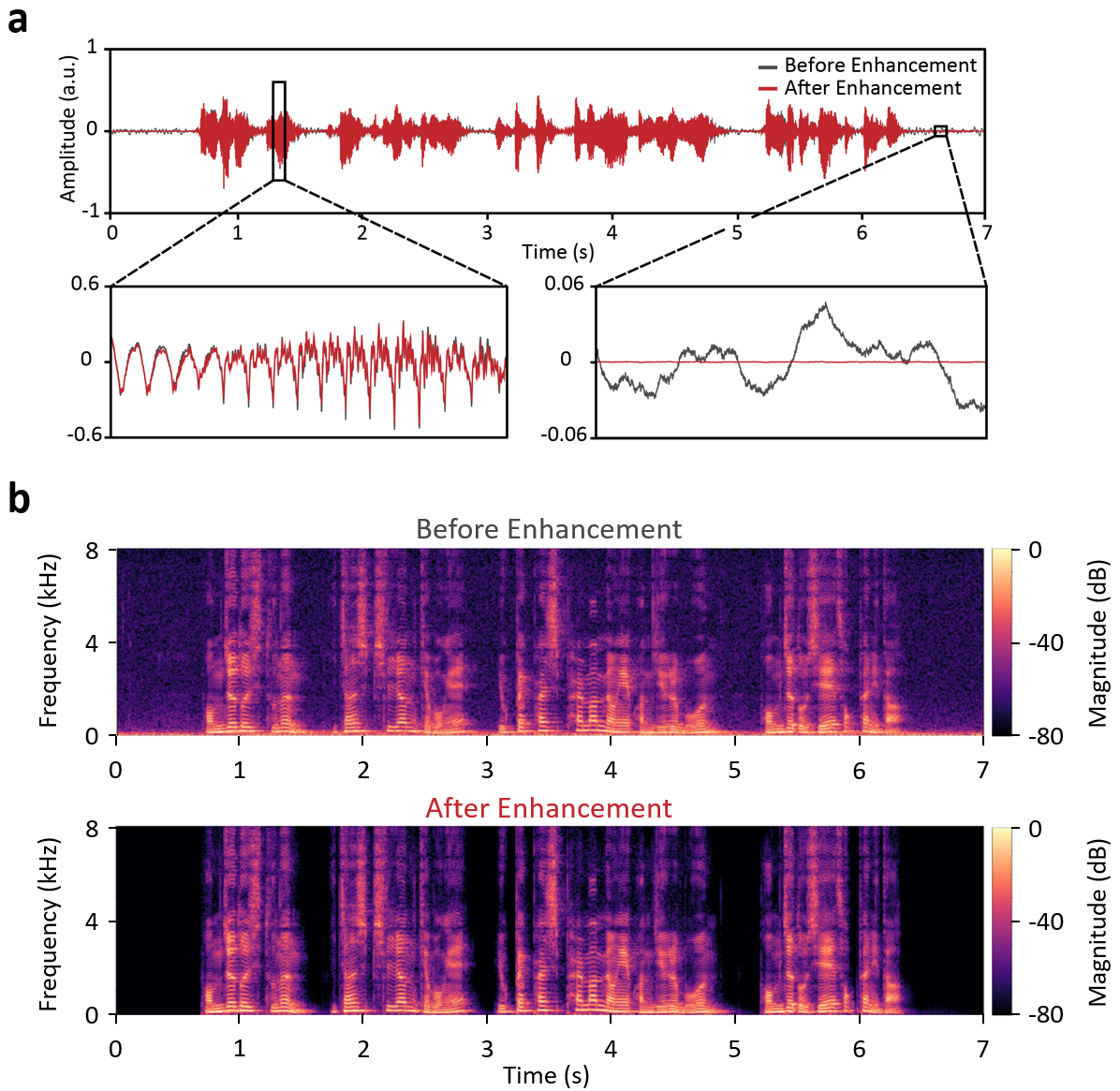}
\caption{\textbf{Noise reduction using a speech enhancement model.} We used the causal version of Demucs\cite{defossez2020real}, pre-trained on the Valentini\cite{Valentini} dataset. (a) Waveforms of the acoustic microphone signal before (black) and after (red) denoising. The lower-left graph provides a close-up of a speech segment, while the lower-right graph zooms in on a noise-only segment. Waveform amplitudes are shown in arbitrary units (a.u.) after amplitude normalization. (b) Spectrograms of the acoustic microphone signal before and after denoising. The Demucs denoising effectively reduces low-level acquisition noise while preserving the speech signal.}
\label{fig:2}
\end{figure}

\subsection*{Recording session}
Utterance scripts were extracted from the Korean newspaper corpus provided by the National Institute of Korean Language (\href{https://corpus.korean.go.kr}{https://corpus.korean.go.kr}). This corpus comprises articles spanning diverse topics such as society, economy, lifestyle, and sports. We curated candidate sentences by applying a length constraint (40–80 characters) and selected 6,000 sentences for recording. Each speaker was randomly assigned 100 unique sentences, with no repetition within a speaker and no overlap across speakers.

The experimental setup for signal recording is shown in Fig. \hyperref[fig:1]{1}c. Measurements were conducted in an acoustically treated recording room at POSTECH. The room is specified to provide high sound isolation (Sound Transmission Class $\geq$ 70), low background noise (Noise Criterion $\leq$ 25--30), and a short reverberation time (Reverberation Time = 0.5--0.7 s at the 500 Hz octave band). Outside noise was not audible to speakers during the recording sessions. The throat microphone was positioned on the supraglottic area of the neck to capture vocal cord vibrations and essential speech formants\cite{song2021study}. The acoustic microphone was placed 30 cm in front of the speaker’s lips. Speakers were instructed to use a forehead rest to maintain a consistent head position throughout the recording session. While speakers read sentences displayed on a screen, speech was recorded simultaneously using both microphones and saved as WAV files on a laptop. During WAV export, the two signals were saved in their raw recorded form without additional adjustment of their relative magnitudes. Recordings were screened via listening and waveform/spectrogram inspection; utterances containing clipping, swallowing, coughing, or other non-speech transients were discarded and re-recorded. To enhance the recording quality, a nylon filter was positioned between the speaker's mouth and the acoustic microphone to prevent pop noise (plosive-induced low-frequency distortion). The nylon filter dissipates the air burst before it reaches the microphone. A reflection filter was also used to minimize ambient noise, and all measurements were conducted using a DC battery to avoid 60 Hz hum interference from the power line.

\subsection*{Post-processing}
Post-processing involved several steps: removal of gravitational acceleration, temporal alignment, acoustic-channel noise reduction, trimming, manual review, and upsampling. Since the throat microphone uses a MEMS accelerometer, the raw signal contains a DC offset from gravitational acceleration. To remove this offset, we applied a 5th-order Butterworth high-pass filter with a 50 Hz cut-off frequency using zero-phase filtering (forward-backward filtering)\cite{SignalProcessing2010}, which applies the filter forward and backward to cancel phase distortion and preserve temporal alignment. Data mismatches due to timing differences between the throat and acoustic microphone measurements were addressed as detailed in the Temporal alignment section. To reduce low-level acquisition noise in the acoustic microphone recordings, we applied Demucs\cite{defossez2020real} (causal version, pre-trained on the Valentini\cite{Valentini} dataset). This step lowers the noise floor, providing a cleaner acoustic target signal for baseline training. Fig. \hyperref[fig:2]{2} illustrates the acoustic microphone signal before and after applying Demucs, demonstrating effective noise reduction without adversely affecting the speech signal. We trimmed leading and trailing non-speech regions using a predefined energy-based criterion. Speech boundaries were detected using an energy threshold set to three times the baseline estimated from silent portions. We then trimmed each recording to the detected speech region with short leading and trailing margins (mean 0.30 s; range 0.15--0.45 s). Each utterance was carefully reviewed to ensure that the recorded speech accurately matched the intended sentence, confirming the correctness of the spoken content. Finally, the throat microphone speech (originally 8 kHz) was upsampled to 16 kHz to match the acoustic microphone's sampling rate. We upsampled the throat microphone signal from 8 kHz to 16 kHz using Fourier-based resampling\cite{SignalProcessing2010}, a standard method for sampling rate conversion that applies an ideal low-pass filter at the original Nyquist frequency (4 kHz), preventing aliasing while preserving the spectral content within the original bandwidth.

\subsection*{Temporal alignment}
\begin{figure}
\centering
\includegraphics[width=\linewidth]{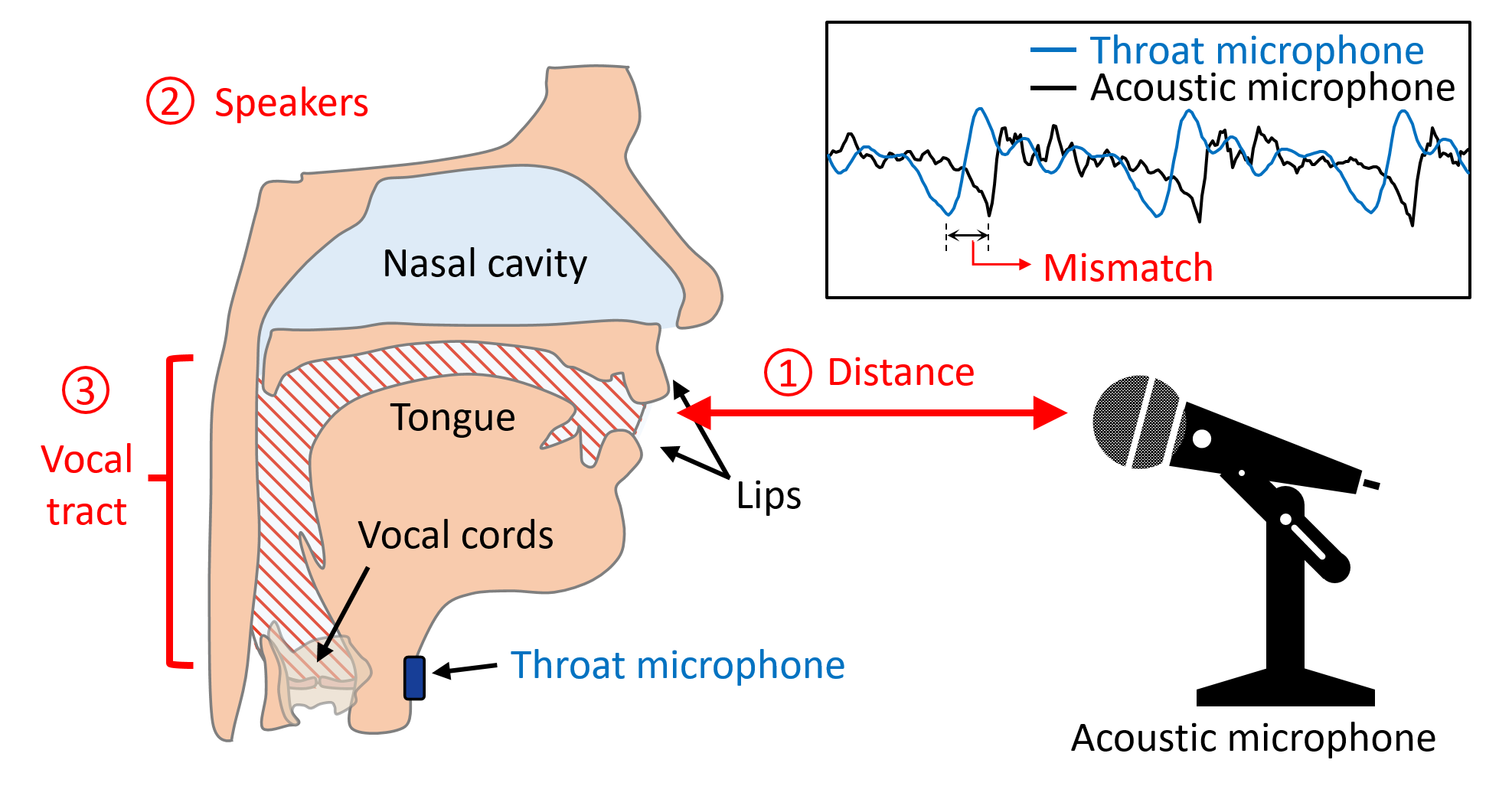}
\caption{\textbf{Factors contributing to timing difference between throat and acoustic microphone signals.} These factors include: (1) the distance between speaker's lips and acoustic microphone; (2) variations in the vocal tract due to changes in the speaker's larynx and oral structure; and (3) changes in the shape of the vocal tract and resonance location depending on the phonemes being produced. The graph in the upper-right corner provides a close-up view of the signal mismatch between the throat and acoustic microphones.}
\label{fig:3}
\end{figure}

We analyzed key factors contributing to temporal misalignment between throat and acoustic microphone signals to enable optimal alignment during post-processing. Previous studies determined mismatches by calculating the sample shift that maximizes the cross-correlation function\cite{Hauret2023configEBEN}. This mismatch value $\delta$ is calculated as follows:
\begin{align}
    \delta=\operatorname{argmax}_k \sum_n T[n]\cdot A[n+k],
\end{align}
where $n$ is the time index, $T[n]$ is the throat microphone signal, $A[n]$ is the acoustic microphone signal. $\delta$ represents the point of strongest linear relationship, indicating optimal temporal alignment. We studied how the mismatch varies depending on the vocalization environment. Fig. \hyperref[fig:3]{3} illustrates the three main causes of timing difference between throat and acoustic microphone signals: (1) the distance between the acoustic microphone and the speaker's lips; (2) variations in speakers' laryngeal and oral structures; and (3) differences in the phonemes being vocalized.

\begin{figure}
\centering
\includegraphics[width=\linewidth]{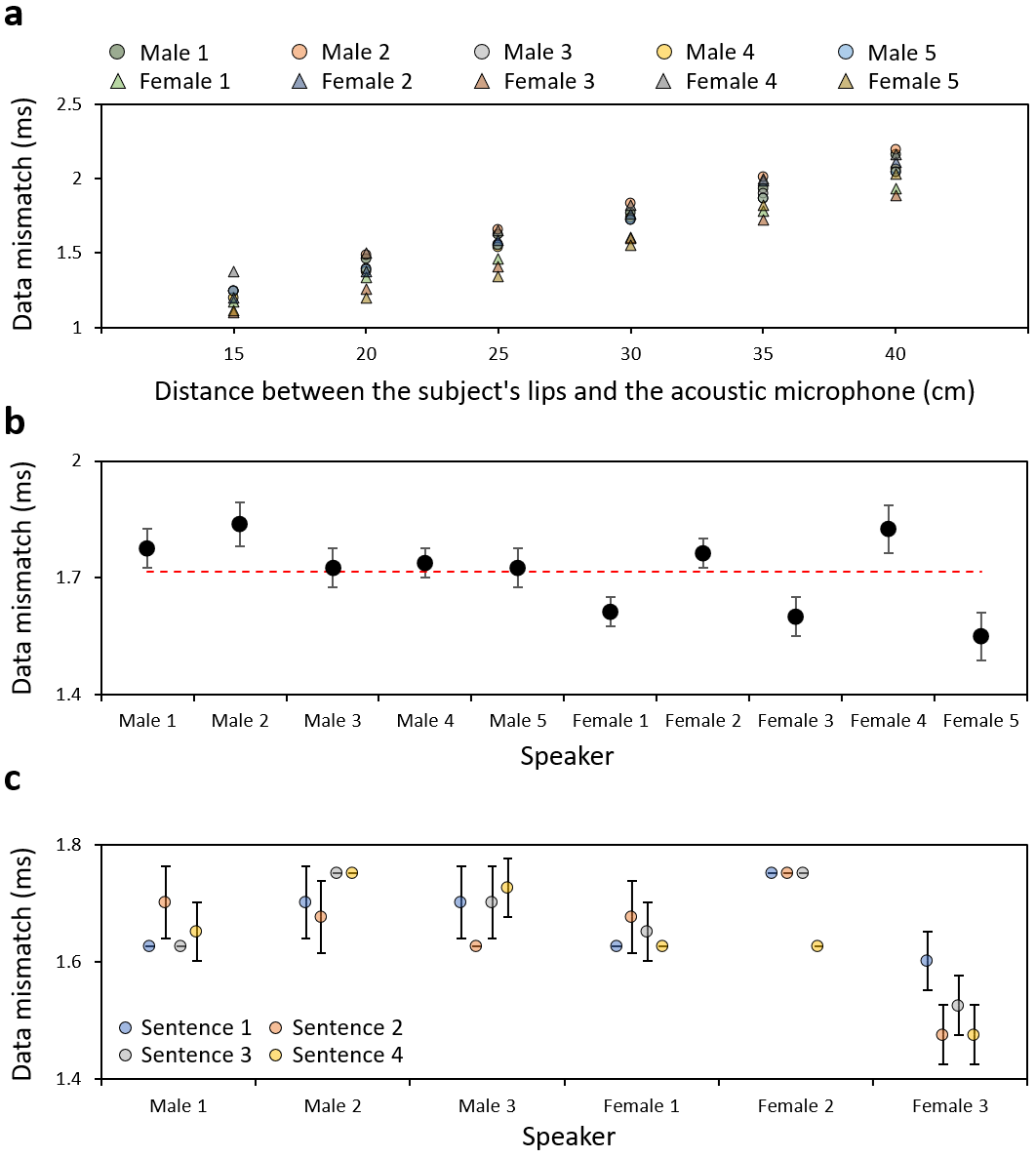}
\caption{\textbf{Analysis of data mismatch between throat and acoustic microphone signals based on the three factors defined in Fig. \hyperref[fig:3]{3}.} Mismatch was calculated using the cross-correlation function between the two signals and is reported in milliseconds (ms). (a) Mismatch as a function of the distance between speaker's lips and acoustic microphone, illustrating a linear increase in mismatch with greater distances. (b) Mismatch among different speakers, highlighting variability when producing the same sentence. The red dashed line represents the mean, and the error bars indicate standard deviations. (c) Mismatch based on different sentences spoken by each speaker, with mismatch variation observed across sentences in all speakers. Error bars represent standard deviations.}
\label{fig:4}
\end{figure}

First, the mismatch increases as the distance between the speaker and the acoustic microphone increases. Fig. \hyperref[fig:4]{4}a shows the calculated mismatch as the distance was varied from 15 to 40 cm; ten speakers (five male, five female) each uttered the same sentence 10 times, and mismatch values were averaged. We varied the distance by sliding the acoustic microphone stand along the table while keeping the speaker’s head position fixed using the forehead rest. A linear increase in mismatch was observed. For clarity, we report mismatch in milliseconds (ms) by converting $\delta$ using the sampling rate. Second, variations in the vocal tract structure among different speakers lead to different timing differences even when they utter the same sentence. Fig. \hyperref[fig:4]{4}b shows the mean and standard deviation of the mismatch, calculated from five male and five female speakers pronouncing the same 10 sentences. The speakers' head positions were fixed, and the distance to the acoustic microphone was kept constant at 30 cm. Despite each speaker uttering the same sentences, the mean mismatch varied. Third, phonemes within a sentence can influence the degree of mismatch, as different sounds resonate in different parts of the vocal tract. For example, some sounds create resonance near the vocal cords, while others resonate in the oral or nasal cavities. This difference affects the arrival time of sound at each microphone and causes timing discrepancies. In particular, voiceless obstruents—sounds made without vocal-cords vibration—can produce little or no signal at the throat microphone, which can lead to variation in the mismatch. Fig. \hyperref[fig:4]{4}c shows the mean and standard deviation of mismatch for each sentence, from recordings of three male and three female speakers, each pronouncing 4 different sentences 5 times. Evidently, the mismatch fluctuates across different sentences.

We considered three main methods to correct these mismatch issues: (1) averaging mismatches over all sentences for a global correction; (2) averaging mismatches for each speaker; and (3) correcting mismatches individually for each sentence. The comparative analysis of these methods' impacts on model performance is presented in the Evaluation of temporal alignment of dataset section.

\begin{figure}
\centering
\includegraphics[width=\linewidth]{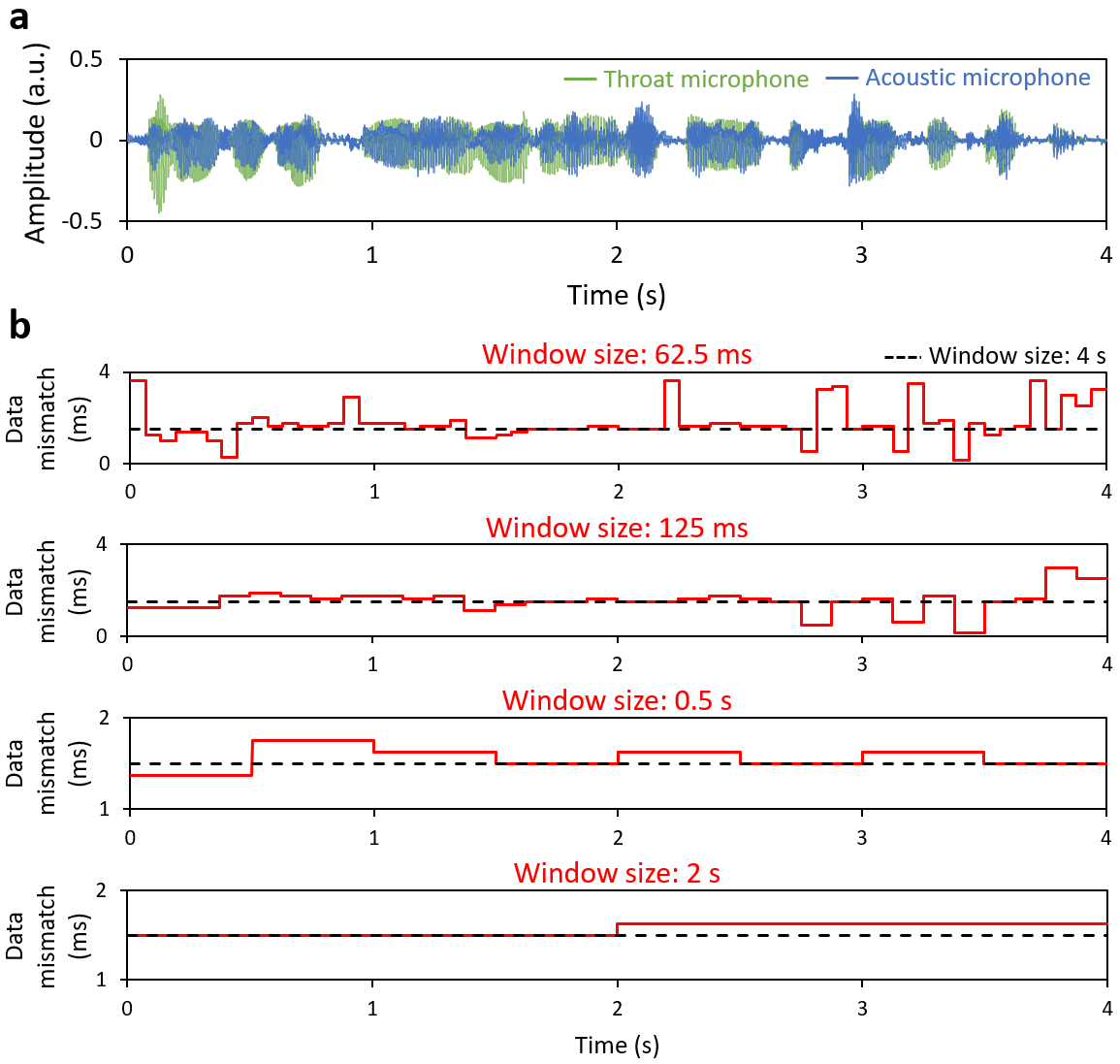}
\caption{\textbf{Analysis of data mismatch between throat and acoustic microphone signals based on window size.} Mismatch was computed using the cross-correlation function over segments of the two signals corresponding to each window size and is reported in milliseconds (ms). (a) Waveforms of simultaneously recorded throat and acoustic microphone signals. Waveform amplitudes are shown in arbitrary units (a.u.) after amplitude normalization. (b) Evaluation of data mismatch across various window sizes. Mismatch variability increases with smaller window sizes, primarily due to non-speech regions and intrinsic differences between throat and acoustic microphone signals.}
\label{fig:5}
\end{figure}

Furthermore, Fig. \hyperref[fig:5]{5} shows the mismatch $\delta$ calculated for each segment by sliding a fixed-size window (0.0625, 0.125, 0.5, and 2 s) over the throat and acoustic microphone signals. When the window size is small (e.g., < 0.125 s), mismatch varies considerably, primarily due to phoneme distribution and silent segments within each window. Conversely, with sufficiently large windows, the impact of these factors on mismatch diminishes.

\subsection*{Ethical declaration}
All procedures in this study were approved by the Institutional Review Board (IRB) at POSTECH (PIRB-2023-E010-R1). We obtained written informed consent from all participants, including consent to share their voice recordings in the released dataset. To protect privacy, participants are represented only by coded speaker IDs (e.g., p00, p01) in the released files and metadata, while direct personal identifiers, such as names and contact information, are excluded. No age information is released, whereas limited metadata such as gender is retained.

\section*{Data Records}
The TAPS dataset is publicly available on the Hugging Face Hub\cite{TAPS} and Zenodo\cite{TAPS_Zenodo}. This corpus contains recordings from 60 native Korean speakers, each producing 100 utterances, simultaneously captured via throat and acoustic microphones. The dataset is split into \texttt{train}, \texttt{dev}, and \texttt{test} sets, with 40 speakers in the \texttt{train} set, 10 in the \texttt{dev} set, and 10 in the \texttt{test} set. Each split maintains a balanced gender distribution. 

On the Hugging Face Hub, the dataset is organized as a \texttt{DatasetDict} with three splits: \texttt{train}, \texttt{dev}, and \texttt{test}. Each split consists of a list of data entries, where each entry corresponds to a specific speaker-utterance pair. Every entry contains the following fields:
\begin{itemize}
    \item \textbf{gender} (\texttt{string}): The speaker's gender: male or female.
    \item \textbf{speaker\_id} (\texttt{string}): A unique identifier for the speaker (e.g., p01).
    \item \textbf{sentence\_id} (\texttt{string}): The utterance index for that speaker (e.g., u30).
    \item \textbf{text} (\texttt{string}): The transcribed sentence. Available for all splits and preserves the original orthography.
    \item \textbf{normalized\_text} (\texttt{string}): A pronunciation-oriented Hangul transcription produced via auditory transcription of the acoustic recordings. Numbers, Latin alphabet sequences, and measurement units are converted into their Korean spoken forms; non-Hangul characters are removed or replaced with spoken equivalents.
    \item \textbf{duration} (\texttt{float32}): The length of the audio recording in seconds.
    \item \textbf{audio.throat\_microphone} (\texttt{Audio}): The throat microphone audio data.
    \item \textbf{audio.acoustic\_microphone} (\texttt{Audio}): The acoustic microphone audio data.
\end{itemize}

Both audio.throat\_microphone and audio.acoustic\_microphone are stored as Hugging Face \texttt{Audio} features, which facilitate on-the-fly audio decoding and easy integration with other datasets and tools in the Hugging Face platform. In particular, each \texttt{Audio} column contains the following sub-fields: 
\begin{itemize}
    \item \textbf{array} (\texttt{array}): The decoded audio data, represented as a 1-dimensional Numpy array.
    \item \textbf{path} (\texttt{string}): The filename of the original WAV file as stored on the Hugging Face Hub. The filename encodes speaker and utterance information in the format \textit{(speaker\_id)\_(sentence\_id)\_(tm/am).wav}, where \textit{tm} and \textit{am} indicate recordings from the throat microphone and acoustic microphone, respectively.
    \item \textbf{sampling\_rate} (\texttt{integer}): The sampling rate of the audio data.
\end{itemize}

Because \texttt{normalized\_text} is derived from auditory transcription of the acoustic recordings, it may differ from \texttt{text}, which preserves the original orthography.

\begin{figure}
\centering
\includegraphics[width=\linewidth]{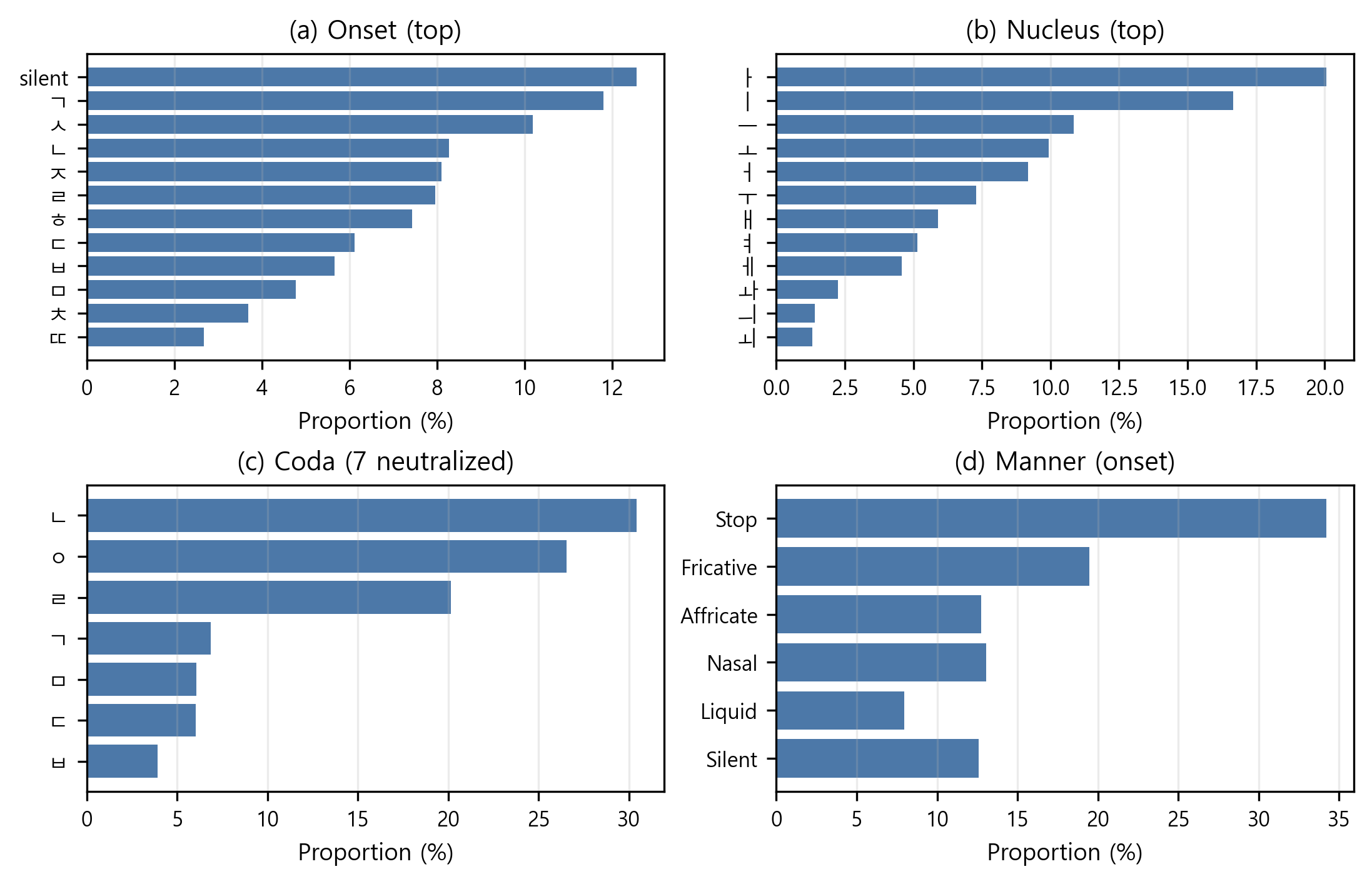}
\caption{\textbf{Phonetic distribution of the TAPS dataset.} (a) Onset (initial consonant) distribution. (b) Nucleus (vowel) distribution. (c) Coda (final consonant, coda7) distribution. (d) Manner-of-articulation distribution. Distributions are consistent across train/dev/test splits.}
\label{fig:6}
\end{figure}

To quantify phonetic coverage, we computed phonetic token distributions from \texttt{normalized\_text}. We applied a Korean grapheme-to-phoneme converter and decomposed Hangul syllables into onset, nucleus, and coda components; codas were further grouped into seven neutralized categories (\textit{coda7}). Across 6,000 utterances, we obtained 290,149 syllables and 692,504 phonetic tokens (consonant-to-vowel ratio: 1.39). Fig. \hyperref[fig:6]{6} summarizes the distributions and shows that they are consistent across splits. In Fig. \hyperref[fig:6]{6}a, \textit{silent} denotes the Korean null onset (orthographic initial <ㅇ>), i.e., vowel-initial syllables with no consonant realization.

In addition to the Hugging Face release (parquet-based), we provide a file-based release on Zenodo in common audio and metadata formats. The Zenodo archive includes paired throat and acoustic microphone recordings as WAV files and per-utterance JSON metadata. Each JSON file contains core metadata (speaker ID, sentence ID, gender, duration, and transcript) and per-channel recording information (sampling rate and number of samples).

\section*{Technical Validation}
This section presents a comprehensive technical validation of the TMSE task, which aims to reconstruct high-quality acoustic microphone signals from throat microphone recordings. Given the limited frequency range and absence of voiceless speech components in throat microphone signals, this task poses significant challenges that require generative modeling approaches capable of inferring missing spectral information. We begin by formulating the training objectives and describing the mathematical models for throat and acoustic signals. To benchmark the task, we evaluated representative baseline models—TSTNN\cite{wang2021tstnn}, Demucs\cite{defossez2020real}, and SE-conformer\cite{kim21seconformer}—that reflect distinct architectural and enhancement strategies. These models were trained and tested on the proposed TAPS dataset, and their performance was assessed using standard speech quality metrics as well as speech intelligibility metrics to verify the restoration of acoustic quality and linguistic content. Through this validation, we demonstrate the feasibility of throat-to-acoustic conversion and highlight architectural choices that contribute to effective speech reconstruction.

\subsection*{Training objectives}
TMSE aims to reconstruct a wideband acoustic signal from a throat microphone input that lacks high-frequency and voiceless components. Because throat microphones are insensitive to airborne turbulence from voiceless obstruents (e.g., /s/ or /f/), models must infer missing high-frequency content from contextual cues in voiced components. This generative requirement distinguishes TMSE from conventional denoising tasks, where target information is present but corrupted by noise.

\subsection*{Baseline models}
\begin{figure}
\centering
\includegraphics[width=\linewidth]{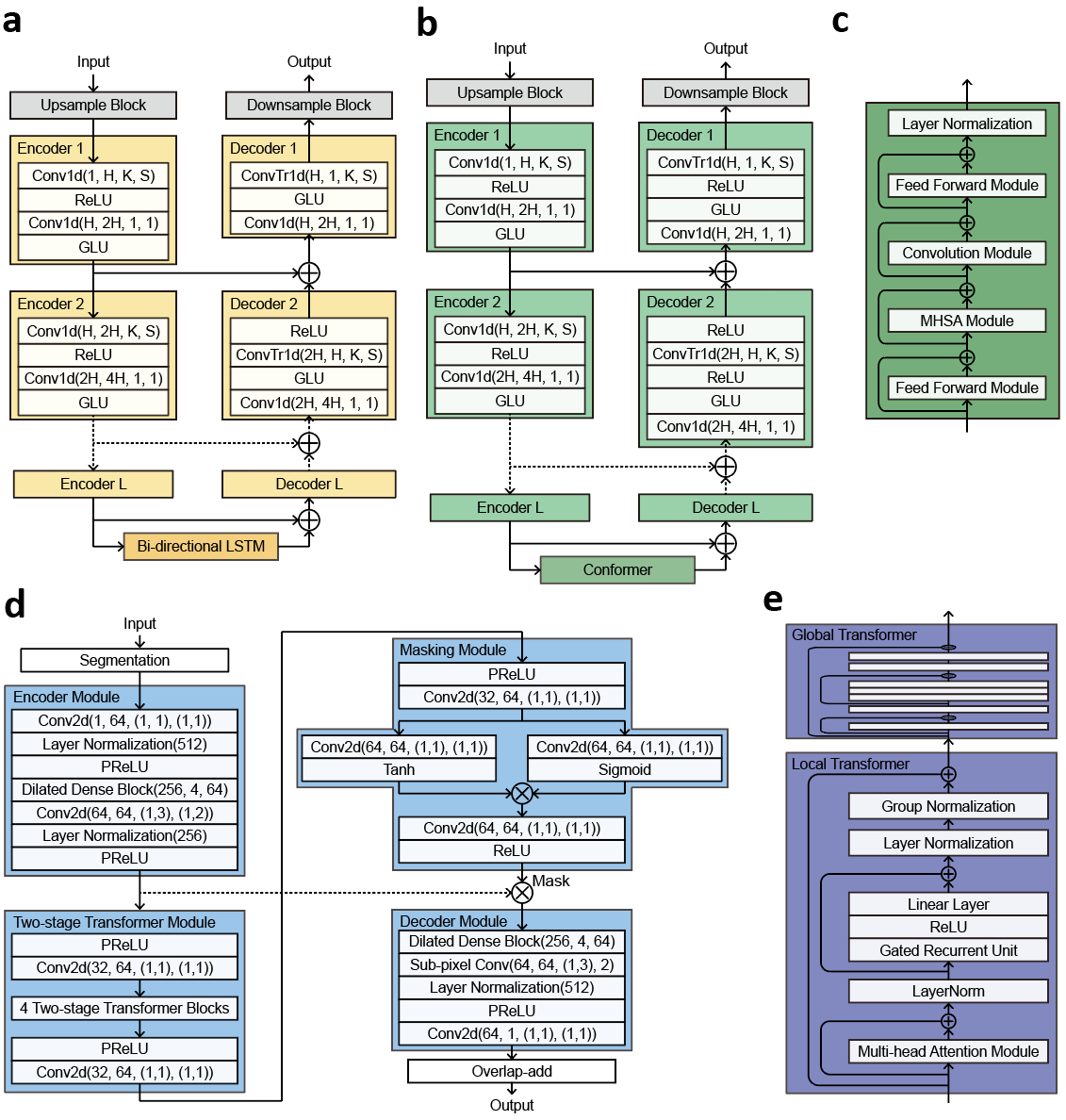}
\caption{\textbf{Block diagram of baseline models.} Architecture of (a) Demucs and (b) SE-conformer. The upsampling factors for the Demucs and SE-conformer are 2 and 4, respectively. The convolution layer parameters follow the format (input channels, output channels, kernel size, stride). For sequence modeling, the Demucs model employs a 2-layer bidirectional long short-term memory, while the SE-conformer model uses a Conformer. (c) Block diagram of the Conformer architecture. (d) Architecture of the TSTNN model. The parameters for the dilated dense block and sub-pixel convolution are presented in the formats (input size, depth, input channels) and (input channels, output channels, kernel size, upsampling rate), respectively. (e) Block diagram of the two-stage transformer block in TSTNN, consisting of two transformer modules.}
\label{fig:7}
\end{figure}

Masking-based approaches modify existing spectral components but cannot generate absent speech elements, whereas mapping-based approaches can infer and reconstruct missing components\cite{Yuliani2021review}. To compare these paradigms for TMSE, we evaluated two mapping-based models (Demucs\cite{defossez2020real} and SE-conformer\cite{kim21seconformer}) and one masking-based model (TSTNN\cite{wang2021tstnn}). \textbf{Demucs} is a convolutional encoder-decoder with U-Net skip connections and BLSTM sequence modeling (Fig. \hyperref[fig:7]{7}a). \textbf{SE-conformer} combines a similar encoder-decoder structure with Conformer blocks for sequence modeling (Fig. \hyperref[fig:7]{7}b,c). \textbf{TSTNN} uses a two-stage transformer module with a multiplicative masking mechanism (Fig. \hyperref[fig:7]{7}d,e). See the original papers for architectural details.

\subsection*{Loss functions}
We followed the loss function implementations from the original papers. For Demucs and SE-conformer, we used an L1 waveform loss combined with a multi-resolution STFT loss (three resolutions)\cite{defossez2020real, kim21seconformer}. For TSTNN, we used an L2 waveform loss combined with a time-frequency domain loss\cite{wang2021tstnn}.

\subsection*{Training configuration}
We configured the Demucs model with the following parameters: kernel size $K=8$, initial hidden channels $H=64$, stride $S=2$, resampling factor $U=2$, and number of layers $L=5$. For the SE-conformer, these parameters were set to $K=4$, $H=64$, $S=4$, $U=4$, and $L=4$. For the Conformer component in the SE-conformer model, we set the input dimension to $512$, the feedforward network dimension to $64$, the number of attention heads to $4$, the depthwise convolution kernel size to $15$, and the Conformer depth to $4$. For TSTNN, we adopted the same implementation as described in the original paper\cite{wang2021tstnn}. We processed the waveform data by segmenting it with a sliding window of 4 seconds and a step size of 2 seconds, resulting in overlapping 4-second segments. From each segment, we randomly selected a starting point between 0 and 2 seconds and extracted a 2-second segment from that point. These 2-second segments were then used for training. All three models were trained for 200 epochs using the Adam optimizer with a learning rate of $3 \times 10^{-4}$, momentum parameters $\beta_1 = 0.9$ and $\beta_2 = 0.99$, and a batch size of 16. For each model, the weights yielding the best performance on the \texttt{dev} set were saved. The final evaluation was conducted on the \texttt{test} set.

\subsection*{Evaluation metrics}
To evaluate the quality of the enhanced speech signal, we employed several objective measures. For speech quality assessment, we used the PESQ\cite{pesq}, specifically the wide-band version recommended in ITU-T P.862.2, which yields scores from -0.5 to 4.5. We also employed the STOI\cite{stoi}, with scores ranging from 0 to 1. Additionally, we adopted three composite measures: CSIG\cite{CsigCbakCovl}, for Mean Opinion Score (MOS) prediction of signal distortion; CBAK\cite{CsigCbakCovl}, for MOS prediction of background noise intrusiveness; and COVL\cite{CsigCbakCovl}, for MOS prediction of the overall effect. All three of these composite scores range from 1 to 5.

To evaluate speech content restoration, we assessed how well the enhanced speech preserved linguistic information, particularly voiceless sounds. For this purpose, we transcribed the enhanced speech using a Whisper-large-v3-turbo ASR model\cite{radford2022whisper}, which was fine-tuned (\href{https://huggingface.co/ghost613/whisper-large-v3-turbo-korean}{https://huggingface.co/ghost613/whisper-large-v3-turbo-korean}) on the Zeroth-Korean dataset (\href{https://openslr.org/40}{https://openslr.org/40}). We then compared these transcriptions with ground-truth labels to compute the character error rate (CER) and word error rate (WER).

\subsection*{Training results}
Table \hyperref[table:2]{2} presents the metric scores for speech quality and content restoration across three baseline models: TSTNN, Demucs, and SE-conformer. Enhancing the throat microphone speech using these models significantly improved overall speech quality, with SE-conformer consistently demonstrating the best performance. Although TSTNN, a masking-based approach, showed competitive objective speech quality scores (PESQ, STOI) compared to the mapping-based Demucs, its speech content restoration quality (CER, WER) was notably lower. These results suggest that mapping-based approaches more effectively restore linguistic content for TMSE, because they can generate missing high-frequency and voiceless components that masking-based methods cannot recover.

\begin{table}
\centering
\caption{Throat microphone speech enhancement results of baseline models. Speech-to-text was performed using Whisper-large-v3-turbo automatic speech recognition model \cite{radford2022whisper}.}
\label{table:2}
\begin{tabular}{>{\centering\arraybackslash}p{4.5cm}|ccccc|cc}
\toprule
Source / Model & PESQ & STOI & \begin{tabular}[c]{@{}c@{}}Predicted\\CSIG\end{tabular} & \begin{tabular}[c]{@{}c@{}}Predicted\\CBAK\end{tabular} & \begin{tabular}[c]{@{}c@{}}Predicted\\COVL\end{tabular} & CER (\%) & WER (\%) \\
\hline
Acoustic Microphone & - & - & - & - & - & 5.5 & 35.3 \\
Throat Microphone & 1.22 & 0.70 & 1.0 & 1.7 & 1.0 & 84.4 & 92.2 \\
\hline
TSTNN, 2021 [\citenum{wang2021tstnn}] & 1.904 & 0.881 & 3.175 & 2.529 & 2.528 & 32.0 & 60.3 \\
Demucs, 2020 [\citenum{defossez2020real}] & 1.793 & 0.883 & 3.177 & 2.442 & 2.470 & 28.7 & 57.4 \\
SE-conformer, 2021 [\citenum{kim21seconformer}] & 1.971 & 0.892 & 3.375 & 2.118 & 2.669 & 24.4 & 53.1 \\
\bottomrule
\end{tabular}
\end{table}

\subsection*{Evaluation of temporal alignment of dataset}

To evaluate the effect of different temporal alignment strategies on model performance, we compared three distinct correction approaches. Let $T_{p,q}[n]$ denote the throat microphone signal, and $A_{p,q}[n]$ denote the acoustic microphone signal for the $q$-th utterance of the $p$-th speaker. 
To determine the temporal shift for any given utterance $(p,q)$, we first identified the time lag $k$ that maximizes the cross-correlation between $T_{p,q}[n]$ and $A_{p,q}[n]$. This process yields an optimal per-utterance shift, which we define as $\delta_{p,q}$:
\begin{equation}
\delta_{p,q} = \operatorname{argmax}_k \sum_n T_{p,q}[n] \cdot A_{p,q}[n+k].
\label{eq:delta_pq}
\end{equation}
The first approach is the \textbf{Per-Utterance Mismatch Correction}, which utilizes the per-utterance shift. In this approach, the individually calculated optimal shift $\delta_{p,q}$ is applied directly to its corresponding utterance $(p,q)$. Thus, the correction value for a given utterance $(p,q)$ is $\Delta_{Utterance}(p,q) = \delta_{p,q}$.

The second approach is the \textbf{Per-Speaker Mean Mismatch Correction}. This approach calculates a unique mean shift for each speaker $p$, denoted as $\Delta_{Speaker}(p)$, by averaging the per-utterance shifts $\delta_{p,q}$ for all $N_{U_p}$ utterances belonging to the speaker:
\begin{equation}
\Delta_{Speaker}(p) = \frac{1}{N_{U_p}} \sum_{q=1}^{N_{U_p}} \delta_{p,q}.
\label{eq:delta_speaker_p}
\end{equation}
This speaker-specific value $\Delta_{Speaker}(p)$ is then applied as a uniform correction to all utterances from speaker $p$.

The final approach is \textbf{Overall Mean Mismatch Correction}. This approach applies a single, global correction value, $\Delta_{Global}$, to all utterances in the dataset. This global value is derived by averaging the per-speaker mean shifts $\Delta_{Speaker}(p)$ across all $N_S$ speakers:
\begin{equation}
\Delta_{Global} = \frac{1}{N_S} \sum_{p=1}^{N_S} \Delta_{Speaker}(p).
\label{eq:delta_global}
\end{equation}

The baseline models were trained using datasets processed with each of these three correction methods. Table \hyperref[table:3]{3} summarizes the percentage differences in PESQ, STOI, and CER for models trained with these methods, relative to training on the uncorrected data. The findings indicate that the overall mean mismatch correction strategy generally yields the most consistent and beneficial impact on performance. For the TSTNN and SE-conformer models, this method achieved enhancements in both PESQ and STOI scores, accompanied by significant reductions in CER. While the Demucs model showed a slight decrease in PESQ with this global averaging approach, it still benefited from a reduction in CER. Considering the balance of improvements across different models and metrics, the overall mean mismatch correction strategy demonstrated robust advantages. The overall mean mismatch correction strategy was adopted for the final TAPS dataset preparation and for all main training procedures reported in this paper. The final applied global shift ($\Delta_{Global}$) is 13 samples at the original 8 kHz sampling rate of the throat microphone (1.625 ms) data.

\begin{table}
\centering
\caption{Percentage difference in objective speech quality metrics (PESQ, STOI, and CER) between different temporal alignment methods, relative to models trained on the uncorrected data. Higher PESQ and STOI values indicate better quality, while lower CER values are preferable.}
\label{table:3}
\begin{tabular}{>{\centering\arraybackslash}c|ccc|ccc|ccc}
\toprule
 & \multicolumn{3}{c|}{ \begin{tabular}[c]{@{}c@{}} Overall Mean\\Mismatch Correction (\%)\end{tabular}} & \multicolumn{3}{c|}{ \begin{tabular}[c]{@{}c@{}} Per-Speaker Mean\\Mismatch Correction (\%)\end{tabular}} & \multicolumn{3}{c}{ \begin{tabular}[c]{@{}c@{}} Per-Utterance\\Mismatch Correction (\%)\end{tabular}}  \\
\midrule
Model & PESQ & STOI & CER & PESQ & STOI & CER & PESQ & STOI & CER \\
\midrule
TSTNN, 2021 [\citenum{wang2021tstnn}]           & 2.33 & 0.88 & $-8.04$ & $-3.27$ & $-0.55$ & $6.97$ & $-1.45$ & $-0.25$ & $2.48$ \\
Demucs, 2020 [\citenum{defossez2020real}]       & $-0.03$ & 0.27 & $-1.24$ & $-1.26$ & $-0.14$ & $2.00$ & $-1.00$ & $-0.02$ & $5.87$ \\
SE-conformer, 2021 [\citenum{kim21seconformer}] & 0.71 & 0.45 & $-4.94$ & $-2.23$ & $-0.30$ & $5.69$ & $-4.98$ & $-0.91$ & $3.75$ \\
\bottomrule
\end{tabular}
\end{table}

\section*{Usage Notes}
The TAPS dataset offers a valuable resource for researchers developing deep learning-based speech enhancement models for throat microphone applications. By providing paired throat and acoustic microphone recordings from a diverse group of native Korean speakers, the TAPS dataset addresses the unique challenges of throat microphone data, such as the loss of high-frequency components. Researchers can leverage this dataset to train models capable of enhancing throat microphone recordings, thereby improving speech intelligibility in noisy environments. The dataset also includes baseline performance metrics from established models like Demucs, SE-conformer, and TSTNN, highlighting the strengths of mapping-based approaches for generating voiceless sounds. Additionally, the dataset introduces standardized methods for temporal alignment, which significantly enhance model accuracy and stability. This resource sets a foundational standard for further exploration and cross-comparative studies in TMSE, providing a pathway for advancements in wearable, noise-resistant communication technologies.

\section*{Data Availability}
The TAPS dataset is available on the Hugging Face Hub (\href{https://huggingface.co/datasets/yskim3271/Throat\_and\_Acoustic\_Pairing\_Speech\_Dataset}{\textnormal{\nolinkurl{https://huggingface.co/datasets/yskim3271/Throat\_and\_Acoustic\_Pairing\_Speech\_Dataset}}}) and Zenodo (\href{https://zenodo.org/records/18324208}{\textnormal{\nolinkurl{https://zenodo.org/records/18324208}}}).

\section*{Code Availability}
The hardware design files and firmware for the recording device are available at \href{https://github.com/yhsong06/taps_hardware}{\textnormal{\nolinkurl{https://github.com/yhsong06/taps_hardware}}}. The preprocessing and temporal-alignment scripts used in the dataset preparation, as well as the code for training and evaluating the baseline models presented in this study, are available at \href{https://github.com/yskim3271/taps-baselines}{\textnormal{\nolinkurl{https://github.com/yskim3271/taps-baselines}}}. The project website is at \href{http://taps.postech.ac.kr}{\textnormal{\nolinkurl{http://taps.postech.ac.kr}}}.

%\bibliography{sample}

\begin{thebibliography}{99}
\urlstyle{same} 

\bibitem{lee2013highly}
Lee, J.-H. \textit{et al.} Highly sensitive stretchable transparent piezoelectric nanogenerators.  
\textit{Energy Environ. Sci.} \textbf{6}, 169--175, \href{https://doi.org/10.1039/C2EE23530G}{\textnormal{\nolinkurl{https://doi.org/10.1039/C2EE23530G}}} (2013).

\bibitem{dagdeviren2014conformable}
Dagdeviren, C. \textit{et al.} Conformable amplified lead zirconate titanate sensors with enhanced piezoelectric response for cutaneous pressure monitoring.  
\textit{Nat. Commun.} \textbf{5}, 4496, \href{https://doi.org/10.1038/ncomms5496}{\textnormal{\nolinkurl{https://doi.org/10.1038/ncomms5496}}} (2014).

\bibitem{park2015fingertip}
Park, J., Kim, M., Lee, Y., Lee, H. S. \& Ko, H. Fingertip skin-inspired microstructured ferroelectric skins discriminate static/dynamic pressure and temperature stimuli.  
\textit{Sci. Adv.} \textbf{1}, e1500661, \href{https://doi.org/10.1126/sciadv.1500661}{\textnormal{\nolinkurl{https://doi.org/10.1126/sciadv.1500661}}} (2015).

\bibitem{kim2016body}
Kim, D. \textit{et al.} Body-attachable and stretchable multisensors integrated with wirelessly rechargeable energy storage devices.  
\textit{Adv. Mater.} \textbf{28}, 748--756, \href{https://doi.org/10.1002/adma.201504335}{\textnormal{\nolinkurl{https://doi.org/10.1002/adma.201504335}}} (2015).

\bibitem{park2016dramatically}
Park, B. \textit{et al.} Dramatically enhanced mechanosensitivity and signal-to-noise ratio of nanoscale crack-based sensors: Effect of crack depth.  
\textit{Adv. Mater.} \textbf{28}, 8130--8137, \href{https://doi.org/10.1002/adma.201602425}{\textnormal{\nolinkurl{https://doi.org/10.1002/adma.201602425}}} (2016).

\bibitem{qiu2015ultrafast}
Qiu, L. \textit{et al.} Ultrafast dynamic piezoresistive response of graphene-based cellular elastomers.  
\textit{Adv. Mater.} \textbf{28}, 194--200, \href{https://doi.org/10.1002/adma.201503957}{\textnormal{\nolinkurl{https://doi.org/10.1002/adma.201503957}}} (2015).

\bibitem{zang2015flexible}
Zang, Y. \textit{et al.} Flexible suspended gate organic thin-film transistors for ultra-sensitive pressure detection.  
\textit{Nat. Commun.} \textbf{6}, 6269, \href{https://doi.org/10.1038/ncomms7269}{\textnormal{\nolinkurl{https://doi.org/10.1038/ncomms7269}}} (2015).

\bibitem{jin2017ultrasensitive}
Jin, M. L. \textit{et al.} An ultrasensitive, visco-poroelastic artificial mechanotransducer skin inspired by piezo2 protein in mammalian Merkel cells.  
\textit{Adv. Mater.} \textbf{29}, 1605973, \href{https://doi.org/10.1002/adma.201605973}{\textnormal{\nolinkurl{https://doi.org/10.1002/adma.201605973}}} (2017).

\bibitem{lee2019ultrathin}
Lee, S. \textit{et al.} An ultrathin conformable vibration-responsive electronic skin for quantitative vocal recognition.  
\textit{Nat. Commun.} \textbf{10}, 2468, \href{https://doi.org/10.1038/s41467-019-10465-w}{\textnormal{\nolinkurl{https://doi.org/10.1038/s41467-019-10465-w}}} (2019).

\bibitem{fan2015ultrathin}
Fan, X. \textit{et al.} Ultrathin, rollable, paper-based triboelectric nanogenerator for acoustic energy harvesting and self-powered sound recording.  
\textit{ACS Nano} \textbf{9}, 4236--4243, \href{https://doi.org/10.1021/acsnano.5b00618}{\textnormal{\nolinkurl{https://doi.org/10.1021/acsnano.5b00618}}} (2015).

\bibitem{yang2015eardrum}
Yang, J. \textit{et al.} Eardrum-inspired active sensors for self-powered cardiovascular system characterization and throat-attached anti-interference voice recognition.  
\textit{Adv. Mater.} \textbf{27}, 1316--1326, \href{https://doi.org/10.1002/adma.201404794}{\textnormal{\nolinkurl{https://doi.org/10.1002/adma.201404794}}} (2015).

\bibitem{kang2018transparent}
Kang, S. \textit{et al.} Transparent and conductive nanomembranes with orthogonal silver nanowire arrays for skin-attachable loudspeakers and microphones.  
\textit{Sci. Adv.} \textbf{4}, eaas8772, \href{https://doi.org/10.1126/sciadv.aas8772}{\textnormal{\nolinkurl{https://doi.org/10.1126/sciadv.aas8772}}} (2018).

\bibitem{zhao2020fully}
Zhao, Y. \textit{et al.} Fully flexible electromagnetic vibration sensors with annular field confinement origami magnetic membranes.  
\textit{Adv. Funct. Mater.} \textbf{30}, 2001553, \href{https://doi.org/10.1002/adfm.202001553}{\textnormal{\nolinkurl{https://doi.org/10.1002/adfm.202001553}}} (2020).

\bibitem{gao2022comparison}
Gao, S. \textit{et al.} Comparison of enhancement techniques based on neural networks for attenuated voice signal captured by flexible vibration sensors on throats.  
\textit{Nanotechnol. Precis. Eng.} \textbf{5}, 013001, \href{https://doi.org/10.1063/10.0009187}{\textnormal{\nolinkurl{https://doi.org/10.1063/10.0009187}}} (2022).

\bibitem{zheng2022dual}
Zheng, C. \textit{et al.} Dual-path transformer-based network with equalization-generation components prediction for flexible vibrational sensor speech enhancement in the time domain.  
\textit{J. Acoust. Soc. Am.} \textbf{151}, 2814--2825, \href{https://doi.org/10.1121/10.0010316}{\textnormal{\nolinkurl{https://doi.org/10.1121/10.0010316}}} (2022).

\bibitem{song2025multimodal}
Song, Y., Yun, I., Giovanoli, S., Easthope, C. A. \& Chung, Y. Multimodal deep ensemble classification system with wearable vibration sensor for detecting throat-related events.
\textit{npj Digit. Med.} \textbf{8}, 14, \href{https://doi.org/10.1038/s41746-024-01417-w}{\textnormal{\nolinkurl{https://doi.org/10.1038/s41746-024-01417-w}}} (2025).

\bibitem{shin2012survey}
Shin, H. S., Kang, H.-G. \& Fingscheidt, T. Survey of speech enhancement supported by a bone conduction microphone.
In \textit{Proc. Speech Commun. 10. ITG Symp.}, 1--4 (2012).

\bibitem{AckerMills2005}
Acker-Mills, B., Houtsma, A. \& Ahroon, W. Speech intelligibility with acoustic and contact microphones. In \textit{Proc. New Dir. Improv. Audio Effect.}, RTO-MP-HFM-123, 7-1--7-14 (2005).

\bibitem{tran2013effect}
Tran, P. K., Letowski, T. R. \& McBride, M. E. The effect of bone conduction microphone placement on intensity and spectrum of transmitted speech items.  
\textit{J. Acoust. Soc. Am.} \textbf{133}, 3900--3908, \href{https://doi.org/10.1121/1.4803870}{\textnormal{\nolinkurl{https://doi.org/10.1121/1.4803870}}} (2013).

\bibitem{toda2012statistical}
Toda, T., Nakagiri, M. \& Shikano, K. Statistical voice conversion techniques for body-conducted unvoiced speech enhancement.  
\textit{IEEE Trans. Audio Speech Lang. Process.} \textbf{20}, 2505--2517, \href{https://doi.org/10.1109/TASL.2012.2205241}{\textnormal{\nolinkurl{https://doi.org/10.1109/TASL.2012.2205241}}} (2012).

\bibitem{mcbride2011effect}
McBride, M., Tran, P., Letowski, T. \& Patrick, R. The effect of bone conduction microphone locations on speech intelligibility and sound quality.  
\textit{Appl. Ergon.} \textbf{42}, 495--502, \href{https://doi.org/10.1016/j.apergo.2010.09.004}{\textnormal{\nolinkurl{https://doi.org/10.1016/j.apergo.2010.09.004}}} (2011).

\bibitem{song2021study}
Song, Y. \textit{et al.} Study on optimal position and covering pressure of wearable neck microphone for continuous voice monitoring.  
In \textit{Proc. 43rd Annu. Int. Conf. IEEE Eng. Med. Biol. Soc.}, 7340--7343, \href{https://doi.org/10.1109/EMBC46164.2021.9629724}{\textnormal{\nolinkurl{https://doi.org/10.1109/EMBC46164.2021.9629724}}} (2021).

\bibitem{vu2007blind}
Vu, T. T., Unoki, M. \& Akagi, M. A blind restoration model for bone-conducted speech based on a linear prediction scheme.  
In \textit{Proc. Int. Symp. Nonlinear Theory Appl.} \textbf{41}, 449--452, \href{https://doi.org/10.34385/proc.41.19AM2-C-5}{\textnormal{\nolinkurl{https://doi.org/10.34385/proc.41.19AM2-C-5}}} (2007).

\bibitem{rahman2017lp}
Rahman, M. A., Shimamura, T. \& Makinae, H. LP-based quality improvement of noisy bone conducted speech.  
\textit{IEEJ Trans. Electron. Inf. Syst.} \textbf{137}, 197--198, \href{https://doi.org/10.1541/ieejeiss.137.197}{\textnormal{\nolinkurl{https://doi.org/10.1541/ieejeiss.137.197}}} (2017).

\bibitem{nakagiri2006improving}
Nakagiri, M., Toda, T., Kashioka, H. \& Shikano, K. Improving body transmitted unvoiced speech with statistical voice conversion.  
In \textit{Proc. Interspeech}, 2270--2273, \href{https://doi.org/10.21437/Interspeech.2006-583}{\textnormal{\nolinkurl{https://doi.org/10.21437/Interspeech.2006-583}}} (2006).

\bibitem{turan2015source}
Turan, M. A. T. \& Erzin, E. Source and filter estimation for throat-microphone speech enhancement.  
\textit{IEEE/ACM Trans. Audio Speech Lang. Process.} \textbf{24}, 265--275, \href{https://doi.org/10.1109/TASLP.2015.2499040}{\textnormal{\nolinkurl{https://doi.org/10.1109/TASLP.2015.2499040}}} (2015).

\bibitem{huang2017wearable}
Huang, B., Gong, Y., Sun, J. \& Shen, Y. A wearable bone-conducted speech enhancement system for strong background noises.  
In \textit{Proc. 18th Int. Conf. Electron. Packag. Technol.}, 1682--1684, \href{https://doi.org/10.1109/ICEPT.2017.8046759}{\textnormal{\nolinkurl{https://doi.org/10.1109/ICEPT.2017.8046759}}} (2017).

\bibitem{liu2018bone}
Liu, H.-P., Tsao, Y. \& Fuh, C.-S. Bone-conducted speech enhancement using deep denoising autoencoder.  
\textit{Speech Commun.} \textbf{104}, 106--112, \href{https://doi.org/10.1016/j.specom.2018.06.002}{\textnormal{\nolinkurl{https://doi.org/10.1016/j.specom.2018.06.002}}} (2018).

\bibitem{zheng2018novel}
Zheng, C., Zhang, X., Sun, M., Yang, J. \& Xing, Y. A novel throat microphone speech enhancement framework based on deep BLSTM recurrent neural networks.  
In \textit{Proc. IEEE 4th Int. Conf. Comput. Commun.}, 1258--1262, \href{https://doi.org/10.1109/CompComm.2018.8780872}{\textnormal{\nolinkurl{https://doi.org/10.1109/CompComm.2018.8780872}}} (2018).

\bibitem{ABCS}
Wang, M., Chen, J., Zhang, X. L. \& Rahardja, S. End-to-end multi-modal speech recognition on an air and bone conducted speech corpus.  
\textit{IEEE/ACM Trans. Audio Speech Lang. Process.} \textbf{31}, 513--524, \href{https://doi.org/10.1109/TASLP.2022.3224305}{\textnormal{\nolinkurl{https://doi.org/10.1109/TASLP.2022.3224305}}} (2022).

\bibitem{Vibravox}
Hauret, J. \textit{et al.} Vibravox: A dataset of French speech captured with body-conduction audio sensors.
\textit{Speech Comm.} \textbf{172}, 103238, \href{https://doi.org/10.1016/j.specom.2025.103238}{\textnormal{\nolinkurl{https://doi.org/10.1016/j.specom.2025.103238}}} (2025).

\bibitem{Vibravox_data}
Hauret, J. \textit{et al.} Vibravox. \textit{Hugging Face}, \href{https://doi.org/10.57967/hf/2727}{\textnormal{\nolinkurl{https://doi.org/10.57967/hf/2727}}} (2024).

\bibitem{wang2021tstnn}
Wang, K., He, B. \& Zhu, W.-P. TSTNN: Two-stage transformer-based neural network for speech enhancement in the time domain.  
In \textit{Proc. IEEE Int. Conf. Acoust. Speech Signal Process.}, 7098--7102, \href{https://doi.org/10.1109/ICASSP39728.2021.9413740}{\textnormal{\nolinkurl{https://doi.org/10.1109/ICASSP39728.2021.9413740}}} (2021).

\bibitem{defossez2020real}
Defossez, A., Synnaeve, G. \& Adi, Y. Real time speech enhancement in the waveform domain.  
In \textit{Proc. Interspeech}, 3291--3295, \href{https://doi.org/10.21437/Interspeech.2020-2409}{\textnormal{\nolinkurl{https://doi.org/10.21437/Interspeech.2020-2409}}} (2020).

\bibitem{kim21seconformer}
Kim, E. \& Seo, H. SE-conformer: Time-domain speech enhancement using conformer.  
In \textit{Proc. Interspeech}, 2736--2740, \href{https://doi.org/10.21437/Interspeech.2021-2207}{\textnormal{\nolinkurl{https://doi.org/10.21437/Interspeech.2021-2207}}} (2021).

\bibitem{Speech2024Kwon}
Kwon, J., Hwang, J., Sung, J. E. \& Im, C.-H. Speech synthesis from three-axis accelerometer signals using conformer-based deep neural network.  
\textit{Comput. Biol. Med.}\textbf{182}, 109090, \href{https://doi.org/10.1016/j.compbiomed.2024.109090}{\textnormal{\nolinkurl{https://doi.org/10.1016/j.compbiomed.2024.109090}}} (2024).

\bibitem{Improving2009Erzin}
Erzin, E. Improving throat microphone speech recognition by joint analysis of throat and acoustic microphone recordings.  
\textit{IEEE Trans. Audio Speech Lang. Process.} \textbf{17}, 1316--1324, \href{https://doi.org/10.1109/TASL.2009.2016733}{\textnormal{\nolinkurl{https://doi.org/10.1109/TASL.2009.2016733}}} (2009).

\bibitem{SignalProcessing2010}
Oppenheim, A. V. \& Schafer, R. W. \textit{Discrete-Time Signal Processing} (3rd Edition, Pearson, 2010).

\bibitem{Valentini}
Valentini-Botinhao, C. Noisy speech database for training speech enhancement algorithms and TTS models. \textit{DataShare}, \href{https://doi.org/10.7488/ds/2117}{\textnormal{\nolinkurl{https://doi.org/10.7488/ds/2117}}} (2017).

\bibitem{Hauret2023configEBEN}
Hauret, J., Joubaud, T., Zimpfer, V. \& Bavu, É. Configurable EBEN: Extreme bandwidth extension network to enhance body-conducted speech capture.  
\textit{IEEE/ACM Trans. Audio Speech Lang. Process.} \textbf{31}, 3499--3512, \href{https://doi.org/10.1109/TASLP.2023.3313433}{\textnormal{\nolinkurl{https://doi.org/10.1109/TASLP.2023.3313433}}} (2023).

\bibitem{TAPS}
Kim, Y., Song, Y. \& Chung, Y. TAPS: Throat and acoustic paired speech dataset.
\textit{Hugging Face}, \href{https://doi.org/10.57967/hf/5804}{\textnormal{\nolinkurl{https://doi.org/10.57967/hf/5804}}} (2025).

\bibitem{TAPS_Zenodo}
Kim, Y., Song, Y. \& Chung, Y. TAPS: Throat and acoustic paired speech dataset (Korean). \textit{Zenodo}, \href{https://doi.org/10.5281/zenodo.18324208}{\textnormal{\nolinkurl{https://doi.org/10.5281/zenodo.18324208}}} (2025).

\bibitem{Yuliani2021review}
Yuliani, A. R., Amri, M. F., Suryawati, E., Ramdan, A. \& Pardede, H. F. Speech enhancement using deep learning methods: A review.  
\textit{J. Elektron. Dan Telekomun.} \textbf{21}, 19--26, \href{https://doi.org/10.14203/jet.v21.19-26}{\textnormal{\nolinkurl{https://doi.org/10.14203/jet.v21.19-26}}} (2021).

\bibitem{pesq}
Rix, A. W., Beerends, J. G., Hollier, M. P. \& Hekstra, A. P. Perceptual evaluation of speech quality (PESQ) – a new method for speech quality assessment of telephone networks and codecs.  
In \textit{Proc. IEEE Int. Conf. Acoust. Speech Signal Process.}, 749--752, \href{https://doi.org/10.1109/ICASSP.2001.941023}{\textnormal{\nolinkurl{https://doi.org/10.1109/ICASSP.2001.941023}}} (2001).

\bibitem{stoi}
Taal, C. H., Hendriks, R. C., Heusdens, R. \& Jensen, J. A short-time objective intelligibility measure for time-frequency weighted noisy speech.  
In \textit{Proc. IEEE Int. Conf. Acoust. Speech Signal Process.}, 4214--4217, \href{https://doi.org/10.1109/ICASSP.2010.5495701}{\textnormal{\nolinkurl{https://doi.org/10.1109/ICASSP.2010.5495701}}} (2010).

\bibitem{CsigCbakCovl}
Hu, Y. \& Loizou, P. C. Evaluation of objective quality measures for speech enhancement.  
\textit{IEEE Trans. Audio Speech Lang. Process.} \textbf{16}, 229--238, \href{https://doi.org/10.1109/TASL.2007.911054}{\textnormal{\nolinkurl{https://doi.org/10.1109/TASL.2007.911054}}} (2008).

\bibitem{radford2022whisper}
Radford, A., Kim, J. W., Xu, T., Brockman, G., McLeavey, C. \& Sutskever, I. Robust speech recognition via large-scale weak supervision. 
In \textit{Proc. Int. Conf. Mach. Learn.}, 28492--28518 (2023).

\end{thebibliography}

\section*{Acknowledgements}
This research was supported by the National Research Foundation grant (RS-2025-00516311), by the Institute of Information \& Communications Technology Planning \& Evaluation (IITP) grant (RS-2019-II191906, Artificial Intelligence Graduate School Program), by the Korea Institute for Advancement of Technology grant (MOTIE, HRD Program for Industrial Innovation, RS-2024-00401466), by the High-Performance Computing Support Project (RQT-25-070138), and by the Regional Innovation System \& Education project (Specialized Industry Scale-UP unit) supported by Gyeongsangbuk-do.

\section*{Author Contributions}
Y.K. and Y.S. contributed equally to this work. Y.K., Y.S., and Y.C. conceptualized the study and established the overall research plan. Y.S. designed and fabricated the custom recording hardware for paired signal acquisition using a microphone and an accelerometer. Y.K. and Y.S. coordinated participant recruitment and conducted the recordings. Y.K., Y.S., and Y.C. conducted the analysis of temporal alignment characteristics in the paired recordings. Y.K. performed signal post-processing and developed the deep learning models. Y.K. and Y.S. drafted the manuscript, and all authors contributed to its revision and final editing. Y.C. supervised the entire project as the corresponding author and secured funding support. All authors have approved the final version of the manuscript.

\section*{Competing Interests}
The authors declare no competing interests.

\end{document}